\documentclass[12pt]{article}

\usepackage[english]{babel}

\usepackage[letterpaper,top=2cm,bottom=2cm,left=3cm,right=3cm,marginparwidth=1.75cm]{geometry}

\usepackage{amsmath}
\usepackage{graphicx,natbib}
\usepackage[colorlinks=true, allcolors=blue]{hyperref}
\usepackage{multirow}
\usepackage{graphics,booktabs}
\usepackage{amsthm,amssymb,authblk}

\theoremstyle{definition}

\usepackage{algorithm,algpseudocode}

\title{On the selection of optimal subdata for big data regression based on leverage scores}
\author{Vasilis Chasiotis}
\author{Dimitris Karlis}

\affil{\small Department of Statistics, Athens University of Economics and Business,  Greece}

\date{}

\usepackage{lipsum}

\begin{document}
	\maketitle
	
	\begin{abstract}
		The demand of computational resources for the modeling process increases as the scale of the datasets does, since traditional approaches for regression involve inverting huge data matrices. The main problem relies on the large data size, and so a standard approach is subsampling that aims at obtaining the most informative portion of the big data. In the current paper, we explore an existing approach based on leverage scores, proposed for subdata selection in linear model discrimination. Our objective is to propose the aforementioned approach for selecting the most informative data points to estimate unknown parameters in both the first-order linear model and a model with interactions. We conclude that the approach based on leverage scores improves existing approaches, providing simulation experiments as well as a real data application.
	\end{abstract}
	
	
	\section{Introduction}
	\label{introduction}
	Due to the size and complexity of big datasets, they can easily exceed the capacity of traditional computing resources. This may lead to the inability to perform certain analyses altogether. Also, more memory or processing power may be required by some statistical models. However, due to unavailability on standard machines, the analysis process can be really complicated.
	
	An approach, in order to address these challenges, is the selection of subdata from the big data to conduct the analysis. Data reduction performs the analysis on a smaller sample that has been selected from the full data. Such an approach leads to reduction of the computational resources that are required for analysis, and so a targeted analysis on the smaller dataset is possible.
	
	The selection of the subdata should be carefully done, in order to ensure that they are representative of the big data, and so the conclusions from the analysis of the subdata can be extrapolated to the big data. Overall, due to computational resource limitations, significant challenges for statistical analyses can be posed by big data, data reduction can be a useful technique for overcoming these challenges. However, before the implementation of the subsampling,  it is really important to take into consideration any potential drawbacks and limitations.
	
	\cite{drineas2011faster} proposed to select randomly a portion of the data. Their idea based on a randomized Hadamard transform on data and then to take subdata at random using uniform subsampling. Their goal was the approximation of the ordinary least-square estimator in linear regression models. However, their approach suffers from the inherent randomness. 
	
	As a consequence, an alternative approach, rapidly developing in recent years, focuses on selecting data points deterministically, so that a small portion of the full data preserves most of the information contained in the full data. Since optimal-design problems relies on data selection, such approaches are connected with the concept of the design of experiments. Therefore, the theory of optimal designs can be very useful in establishing a framework to select the most informative subdata from the full data.
	
	For the remaining of the paper, we will use $n$, $p$ and $k$ to represent the full data size, the number of covariates and the subdata size, respectively.
	
	As a first attempt, \cite{wang2019information} proposed the information-based optimal subdata selection (IBOSS) approach, which is motivated by the concept of optimal experimental designs. They focused on the selection of the most informative subdata from the full data for the estimation of unknown parameters. Their idea was the ``maximization" of an information matrix, that is the central goal in the theory of  optimal experimental designs. Overall, they concluded that subdata that maximizes the determinant of the inverse of the covariance matrix of the unknown parameters is D-optimal, and so it contains the most informative data points from the full data. Also, they developed an algorithm, in order to select D-optimal subdata, based on an upper bound of the determinant of the inverse of the covariance matrix of the unknown parameters. To be more precise, the algorithm of the IBOSS approach selects data points with the smallest as well as largest values of all covariates sequentially, given that previous selected data points are excluded, and its time complexity is $O(np+kp^2)$, or $O(np)$ when $n>kp$.
	
	\cite{wang2021oss} proposed the orthogonal subsampling (OSS) approach to select subdata, that is their approach is based on the optimality of two-level orthogonal arrays. We need to mention that a two-level orthogonal array minimizes the average variance of the estimated parameters as well as provides the best predictions \citep{dey&mukerjee}, and so it represents an optimal design for linear regression. The sequential addition algorithm developed by \cite{wang2021oss} is based on the combinatorial orthogonality of a two-level orthogonal array. Also, they prevent the algorithm to be time-consuming, by eliminating data points, and so its computational complexity is $O(np\text{log}k)$. The algorithm is based on a discrepancy function that measures the distortion of data points on keeping two features simultaneously that are connected with the optimality of orthogonal arrays. The first feature is the selection of extreme data points and the second one is that the signs of the selected data points are as dissimilar as possible (combinatorial orthogonality). Moreover, the OSS approach outperforms the IBOSS approach for the selection of informative subdata from the full data.
	
	The approach of \cite{ren&zhao}, motivated by \cite{wang2021oss}, focuses on selecting subdata that approach a $k\times p$ two-level orthogonal array of strength 2. However, \cite{chasiotis2023} mentioned an issue about implementation of Algorithm 3 of \cite{ren&zhao}, and so their approach is not taken into consideration.
	
	Furthermore, the approach of \cite{chasiotis2023} aims at identifying and interchanging data points that were not selected with those that have already been selected by an approach, i.e. the OSS or the IBOSS one, in order to improve the value of the D-optimality criterion. The two proposed algorithms in \cite{chasiotis2023} are considered as extensions to existing ones, and so they should be evaluated considering a trade-off between improving the value of the D-optimality criterion at the cost of some additional computational time. 
	
	The approach of \cite{wang2019information} has been extended to other cases, e.g. multinomial logistic regression \citep{yao2019optimal}, quantile regression  \citep{wang2021optimal}, and for logistic regression \citep{cheng2020information}. Also, for further related work we refer the reader to \cite{wang2019divide}, \cite{lee2021fast}, and \cite{deldossi2021optimal}. More information on subdata selection or subsampling from big data based on designs can be found in the review papers by \cite{yao2021review} and \cite{yu2023}, which provides a comprehensive overview of the current state of research in this area.
	
	The approach in the current paper, which is a deterministic selection of the most informative data points from the full data based on leverage scores (LEVSS), has already been proposed to select subdata for linear model discrimination by \cite{yu2022}. However, there are some motivations that drive us to further investigate the role of leverage scores in the selection of the most informative data points in order to estimate unknown parameters. At first, since the IBOSS and OSS approaches obtain subdata that are D-optimal, Theorem 2 in \cite{yu2022} motivates us that the selection of the most informative data points based on LEVSS approach, to estimate unknown parameters, should be further investigated. Moreover, we are motivated by \cite{chasiotis2023}, who focused on selecting data points with large convex hull as close as possible to the one generated by the full data, that is, under the subdata, the determinant of the information matrix will be large. \cite{chasiotis2023} also proved that the maximization of the determinant of the information matrix can be addressed as the maximization of the generalized variance of covariates under the selected subdata. It is important to note that the volume of space occupied by the cloud of the selected data points is proportional to the square root of the generalized variance.
	
\begin{figure}[!thb]
		\centering
		\includegraphics[width=1\textwidth]{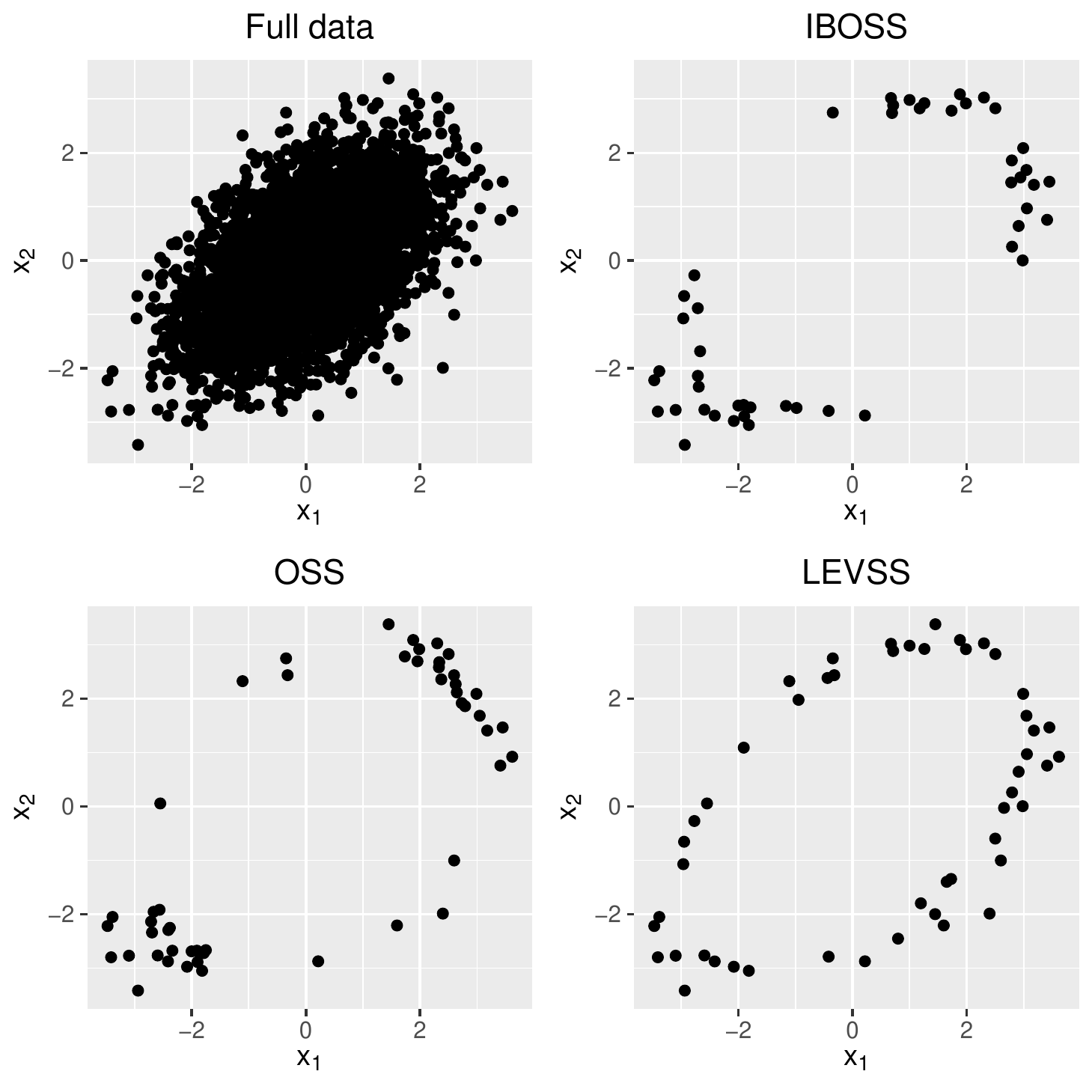}
		\caption{An example for the different approaches. A dataset of $n=5000$ and $p=2$ was generated. The different approaches were used to select $50$ data points. In the first row, the full data can be seen in the first panel and the IBOSS approach in the second one. In the second row, the OSS and LEVSS approaches can be seen in the first and the second panel, respectively.}
		\label{mot}
	\end{figure}

	To provide further information about our motivation, consider the data in Figure \ref{mot}. We have $n=5000$ and $p=2$. Observations $\textbf{x}_1$ and $\textbf{x}_2$ follow a multivariate normal distribution, that is, $\textbf{x}_i\sim N(\textbf{0},\mathbf{\Sigma})$, where $\mathbf{\Sigma}=\left(\Sigma_{ij}\right)$, $i,j=1,2$ is a covariance matrix. Also, $\Sigma_{ij}=1$ for $i=j=1,2$ and $\Sigma_{ij}=0.5$ for $i\ne j=1,2$. Suppose that we are interested in selecting $k=50$ data points. The IBOSS approach selects the extreme data points of the two covariates, and the OSS approach selects data points that are located as close as possible at the corners of the data domain. The LEVSS approach seems to select data points with large convex hull in some sense. It can be seen that the data points selected by the LEVSS approach provide a greater degree of precision about the structural attributes of the full data. Therefore, based on theoretical results of \cite{yu2022} and \cite{chasiotis2023}, the LEVSS approach seems to be really promising for the selection of the most informative data points.
	
	In the review paper by \cite{yu2023}, the authors have evaluated a numerous of algorithms for their ability to select the most informative subdata from big data to estimate unknown parameters, providing simulation experiments as well as a real data application. One of the evaluated algorithms is the algorithm of the LEVSS approach. However, in the current paper we provide further evidence through simulation experiments as well as a real data application, that the LEVSS approach can lead to the selection of the most informative data points in order to estimate unknown parameters. Also, we take into consideration not only the first-order linear model but also a model with interactions. Moreover, we should note that we are interested in comparing the subdata selected based on LEVSS approach with the IBOSS and OSS ones, in case of linear models, that is we assume predictors and responses follow a postulated model. This means that we do not take into consideration model-free subsampling methods, with the aim of elucidating the underlying processes in case of assuming a postulated model.
	
	The remaining of the paper is organized as follows. In Section \ref{theor}, we briefly review the best linear unbiased estimator based on the subdata under a linear regression model. In Section \ref{section_algorithm}, we describe the algorithm of the LEVSS approach. In Section \ref{section_simulation}, we provide simulation evidence to support the LEVSS approach, including a comparison with existing approaches (IBOSS and OSS). In Section \ref{section_real_data}, we use a real dataset for illustration, and in Section \ref{concluding_remarks} this article is concluded with some discussions.
	
	\section{Preliminaries}
	\label{theor}
	Assume that the full data are denoted by ($\textbf{x}_i, y_i$), \ldots, ($\textbf{x}_n, y_n$). Let the linear regression model:
	\begin{equation}\label{model1}
		y_i=\beta_0+\textbf{x}_i^\text{T}\boldsymbol{\beta}_1+\epsilon_i, \quad i=1,2,\ldots,n,
	\end{equation}
	where $\beta_0$ is the intercept parameter, $\textbf{x}_i=(x_{i1},x_{i2},\ldots,x_{ip})^{\text{T}}$ is a covariate vector, $\boldsymbol{\beta}_1=(\beta_1,\beta_2,\ldots,\beta_p)^{\text{T}}$ is a $p$-dimensional vector of unknown slope parameters, $y_i$ is a response, and $\epsilon_i$ is an error term. The $y_i$'s are uncorrelated given the covariates $\textbf{x}_i$, $i=1,2,\ldots,n$ and $\epsilon_i$'s satisfy E$(\epsilon_i)=0$ and V$(\epsilon_i)=\sigma^2$. 
	
	Taking into consideration the full data under model \eqref{model1}, the least-square estimator of $\boldsymbol{\beta}=(\beta_0,\boldsymbol{\beta}_1^{\text{T}})^{\text{T}}$, that is its best linear unbiased estimator, is
	\begin{equation*}
		\hat{\boldsymbol{\beta}}_{\text{Full}}=\left(\sum_{i=1}^{n}\textbf{z}_i\textbf{z}_i^{\text{T}}\right)^{-1}\sum_{i=1}^{n}\textbf{z}_iy_i,
	\end{equation*}
	where $\textbf{z}_i=(1,\textbf{x}_i^\text{T})^\text{T}$.
	
	The inverse of
	\begin{equation*}
		\textbf{Q}_{\text{Full}}=\dfrac{1}{\sigma^2}\sum_{i=1}^{n}\textbf{z}_i\textbf{z}_i^{\text{T}}
	\end{equation*}
	is equal to the covariance matrix of $\hat{\boldsymbol{\beta}}_{\text{Full}}$, where $\textbf{Q}_{\text{Full}}$ is the observed Fisher information matrix of $\boldsymbol{\beta}$ for the full data in case that the error terms $\epsilon_i$'s are normally distributed. $\textbf{Q}_{\text{Full}}$ is still called the information matrix, even though the normality assumption is not required.
	
	If the sample size $n$ of the full data is too large, then a full analysis of the whole data may be infeasible. Therefore, based on limitations of the computational resources, we are interested in gaining useful information from the full data by the selection of a subset of the full data.
	
	Let $\delta_i$ be a indicator variable about the inclusion of ($\textbf{x}_i, y_i$) in the subdata. Therefore, $\delta_i=0$ if ($\textbf{x}_i, y_i$) is not included in the subdata and $\delta_i=1$ otherwise. Also, we assume that we want to select subdata of size $k$, that is $\sum_{i=1}^{n}\delta_i=k$. Thus, the least-square estimator of $\boldsymbol{\beta}$ is still the best linear unbiased estimator based on the subdata, that is, 
	\begin{equation*}
		\hat{\boldsymbol{\beta}}_{\text{Sub}}=\left(\sum_{i=1}^{n}\delta_i\textbf{z}_i\textbf{z}_i^{\text{T}}\right)^{-1}\sum_{i=1}^{n}\delta_i\textbf{z}_iy_i.
	\end{equation*}
	
	The information matrix under the subdata of size $k$ can be written as
	\begin{equation}\label{sub_inf}
		\textbf{Q}_{\text{Sub}}=\dfrac{1}{\sigma^2}\sum_{i=1}^{n}\delta_i\textbf{z}_i\textbf{z}_i^{\text{T}}.
	\end{equation}
	The selected subdata should be optimal is some way. According to the D-optimality criterion, subdata of size $k$ is D-optimal if the determinant of the corresponding $\textbf{Q}_{\text{Sub}}$ is maximized.
	
	Also, we should mention that \cite{chasiotis2023} proved that the determinant of $\textbf{Q}_{\text{Sub}}$ in \ref{sub_inf} is the generalized variance \citep{wilks1932genvar} of covariates $\textbf{x}_i$'s under the selected subdata, and so they addressed the problem of maximizing the determinant of $\textbf{Q}_{\text{Sub}}$ in \ref{sub_inf} as a problem of maximizing the generalized variance of covariates under the selected subdata.
	
	\section{Leverage score based algorithm}\label{section_algorithm}
	In this section, we provide the deterministic leverage score selection (LEVSS) algorithm  proposed by \cite{yu2022} for linear model discrimination.
	
	Following the notations given by \cite{yu2022}, let $\#\left(\mathit{\Gamma}\right)$ denote the cardinal number of a set $\mathit{\Gamma}$, and $\kappa\left(\textbf{B}\right):=\lambda_{max}\left(\textbf{B}\right)/\lambda_{min}\left(\textbf{B}\right)$ denote the condition number of a matrix $\textbf{B}$, where $\lambda_{max}\left(\textbf{B}\right)$ and $\lambda_{min}\left(\textbf{B}\right)$ are the maximum and minimum eigenvalues of the squared matrix $\textbf{B}$, respectively. Also, when $\textbf{B}$ is a singular matrix, then $\kappa\left(\textbf{B}\right)=\infty$.
	
	The LEVSS algorithm is provided in Algorithm \ref{levss}.
	
	\begin{algorithm}
		\caption{LEVSS}
		\label{levss}
		\begin{algorithmic}
			\Require The design matrix $\textbf{X}=(\textbf{x}_{i}^\text{T}), i=1,2,\ldots,n$,  the target sample size ($k>p$), and the threshold $T$ ($\ge1$).
			\Ensure The selected index set $\mathit{\Gamma}$ and the design matrix under the subdata.
			\State \hspace{-0.55cm} \textbf{Initialization:} $\mathit{\Gamma}=\O$, $\textbf{U}_{\mathit{\Gamma}}=\O$, $\kappa\left(\textbf{U}_{\mathit{\Gamma}}^\text{T}\textbf{U}_{\mathit{\Gamma}}\right)=\infty$.
			\State Perform a singular value decomposition of $\textbf{X}$ as $\textbf{X}=\textbf{U}\boldsymbol{\Sigma}\textbf{V}^\text{T}$, calculate the leverage scores $h_{ii}:=\lVert U_{i\cdot}\rVert^2$, where $U_{i\cdot}$ denotes the $i$th row of $\mathbf{U}$, and sort $h_{ii}$'s to have $h_{(11)}\ge\ldots\ge h_{(nn)}$.
			\For{$i$ in $1,\ldots,n$}
			\If{$\#\left(\mathit{\Gamma}\right)\le k$ or $\kappa\left(\textbf{U}_{\mathit{\Gamma}}^\text{T}\textbf{U}_{\mathit{\Gamma}}\right)\ge T$}
			\State Add the index of the data point corresponding to $h_{(ii)}$ to set $\mathit{\Gamma}$.
			\State Update the $\textbf{U}_{\mathit{\Gamma}}$ as the selected rows of $\mathbf{U}$ in $\mathit{\Gamma}$.
			\Else
			\State \textbf{break}
			\EndIf
			\EndFor
		\end{algorithmic}
	\end{algorithm}
	
	\cite{yu2022} mentioned that the stopping criterion for the LEVSS algorithm on the condition number is in order to ensure that the design matrix under the subdata is not ill-conditioned, that is to prevent multicollinearity. Also, they mentioned that, from geometrical perspective, the threshold $T$ on the condition number prevents subdata to lie in a low-rank subspace. Moreover, in case that LEVSS algorithm selects more than $k$ data points, say $k^*$, then they suggested a simple random sampling selecting $k$ out of $k^*$.
	
	Furthermore, they remarked that the stopping criterion for the LEVSS algorithm on the condition number is not crucial when the covariates are from the family of elliptically contoured distributions \citep{fang1990}, since the space of covariates of the subdata expands to the space of covariates of the full data in a quick way.
	
	The time complexity of LEVSS algorithm is $O(np^2)$.
	
	\section{Simulation experiments}\label{section_simulation}
	In this section, we evaluate the performance of LEVSS algorithm based on simulated data, presenting the results of the algorithms of the approaches of IBOSS and OSS as well, in order to make a comparison.
	
	\subsection{First-order linear model}\label{section_first_order}
	Under model \eqref{model1}, covariates $\textbf{x}_i$'s are generated according to the following scenarios.
	\begin{itemize}
		\item Case 1. $\textbf{x}_i$'s are independent and have a multivariate uniform distribution on $[0,1]^p$ with all covariates independent.
		\item Case 2. $\textbf{x}_i$'s have a multivariate normal distribution, that is, $\textbf{x}_i\sim N(\textbf{0},\mathbf{\Sigma})$, with 
		\begin{equation}\label{var_matrix}
			\mathbf{\Sigma}=\left(0.5^{I(i,j)}\right), i,j=1,2,\ldots,p,   
		\end{equation}
		where $I(i,j)=0$ for $i=j=1,2,\ldots,p$ and $I(i,j)=1$ for $i\ne j=1,2,\ldots,p$.
		\item Case 3. $\textbf{x}_i$'s have a truncated multivariate normal distribution on $[-5,5]^p$, that is, $\textbf{x}_i\sim N(\textbf{0},\mathbf{\Sigma})$, with covariance matrix $\mathbf{\Sigma}$ in \eqref{var_matrix}.
	\end{itemize}
	
	The response data are generated from the linear model in \eqref{model1} with the true value of $\boldsymbol{\beta}$ being a $51$ dimensional vector with all elements equal to $1$ and $\sigma^2=9$. We include an intercept, and so $p=50$.
	
	The simulation is repeated $1000$ times and empirical mean squared error (MSE) of the subdata selected by the approaches of IBOSS, OSS and LEVSS are calculated. We estimate the intercept with the adjusted estimator $\hat{\beta}_0=\bar{y}-\bar{\textbf{x}}^{\text{T}}\hat{\boldsymbol{\beta}}_1^{Sub}$ \citep{wang2019information}, where $\bar{y}$ is the mean of the response full data, $\bar{\textbf{x}}$ is the vector of means of all covariates in the full data, and $\hat{\boldsymbol{\beta}}_1^{Sub}$ is the ordinary least-square estimator of $\boldsymbol{\beta}_1^{Sub}$ based on the subdata. Therefore, we consider $(\hat{\beta}_0^{(r)}-\beta_0)^2$ and $||\hat{\boldsymbol{\beta}}_1^{(r)}-\boldsymbol{\beta}_1||^2$ the MSE for intercept and slope estimators in the $r$th repetition, respectively, where $\hat{\beta}_0^{(r)}$ and $\hat{\boldsymbol{\beta}}_1^{(r)}$ are $\hat{\beta}_0$ and $\hat{\boldsymbol{\beta}}_1^{Sub}$ in the $r$th repetition.
	
\begin{figure}[!thb]
		{\centering
			\includegraphics[width=1\textwidth]{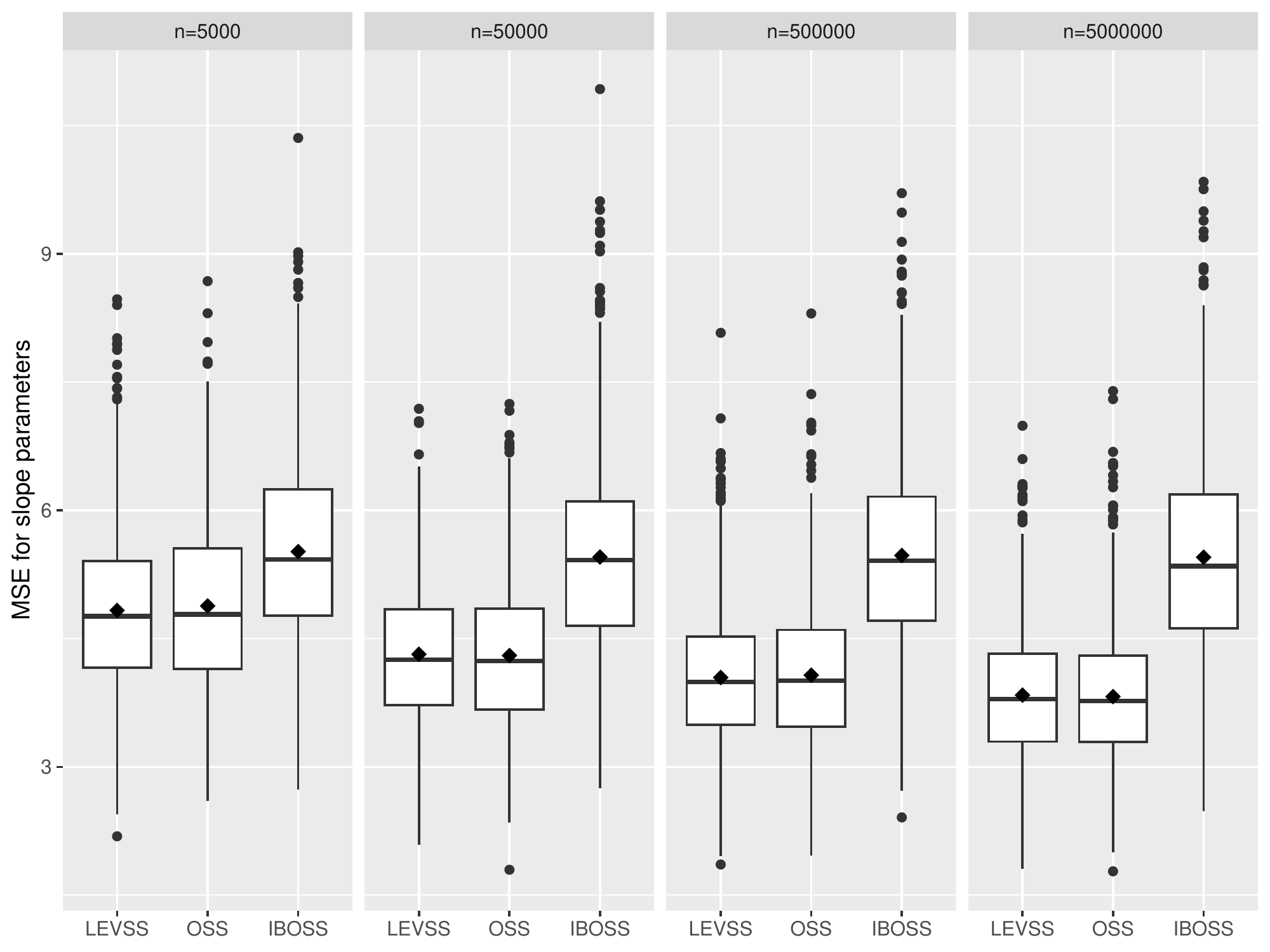}\par}
		\caption{The MSEs of the estimated slope parameters for the subdata selected by different approaches for the covariates of Case 1, when the full data size is $n=5\times10^3, 10^4, 10^5$ and $10^6$ and the subdata size is $k=1000$.}
		\label{fig_uni}
	\end{figure}

\begin{figure}[!thb]
		{\centering
			\includegraphics[width=1\textwidth]{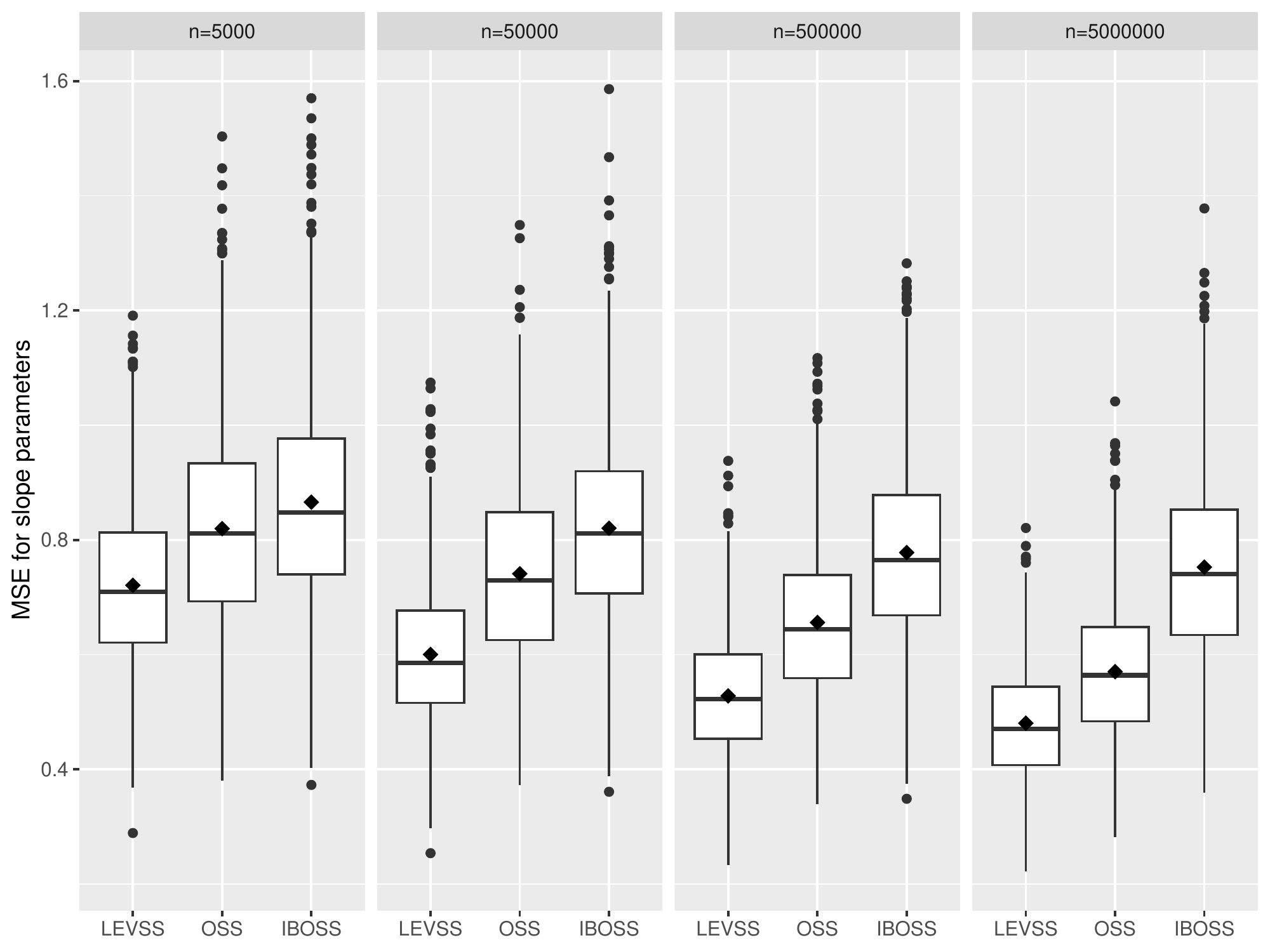}\par}
		\caption{The MSEs of the estimated slope parameters for the subdata selected by different approaches for the covariates of Case 2, when the full data size is $n=5\times10^3, 10^4, 10^5$ and $10^6$ and the subdata size is $k=1000$.}
		\label{fig_norm}
	\end{figure}

\begin{figure}[!thb]
		{\centering
			\includegraphics[width=1\textwidth]{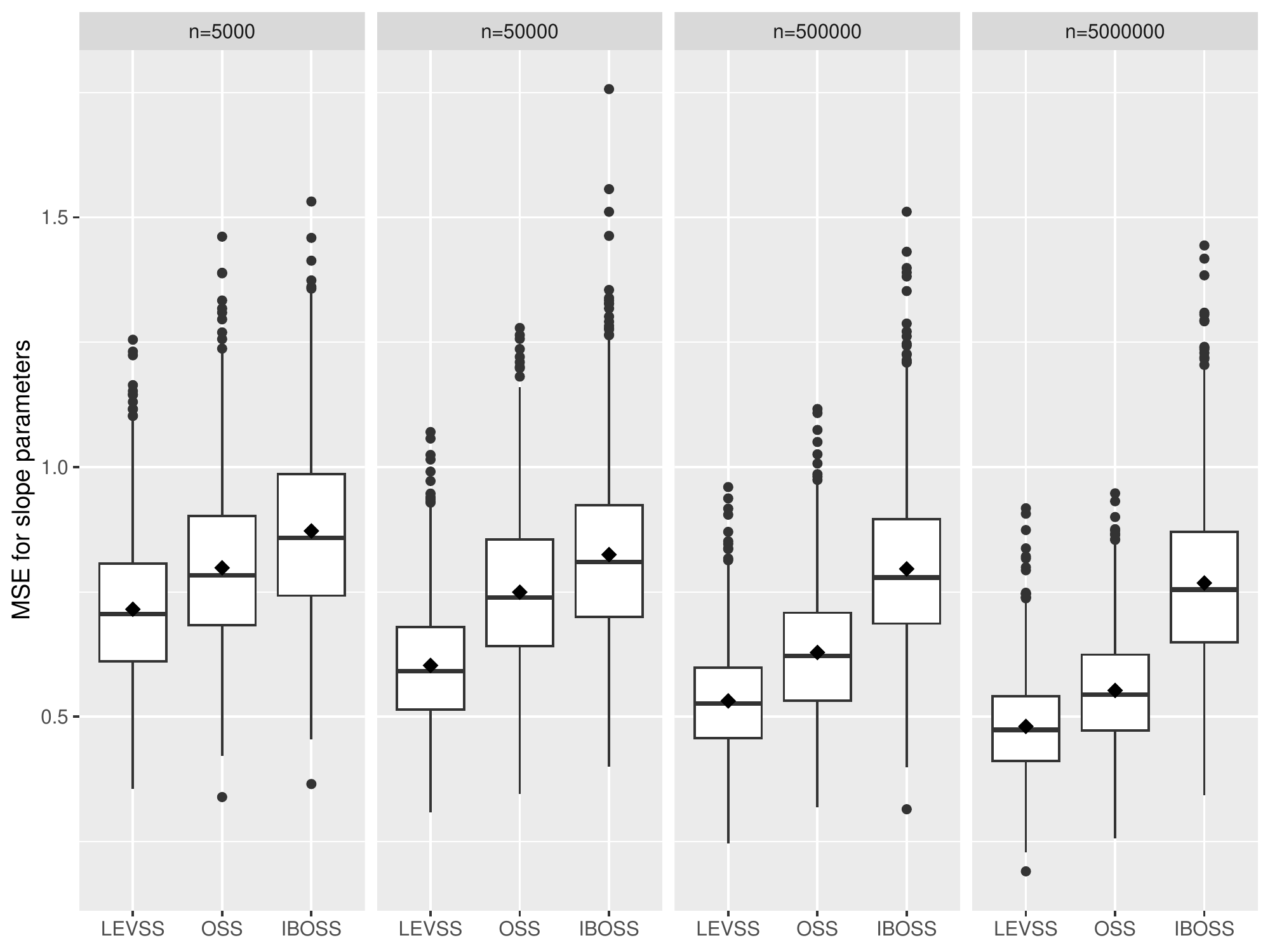}\par}
		\caption{The MSEs of the estimated slope parameters for the subdata selected by different approaches for the covariates of Case 3, when the full data size is $n=5\times10^3, 10^4, 10^5$ and $10^6$ and the subdata size is $k=1000$.}
		\label{fig_tnorm}
	\end{figure}

	We investigate the cases that the full data sizes are $n=5\times10^3, 10^4, 10^5$ and $10^6$, and the subdata size is fixed at $k=1000$. Figures \ref{fig_uni}, \ref{fig_norm} and \ref{fig_tnorm} show the MSEs of the estimated slope parameters for the subdata selected by different approaches for Cases 1, 2 and 3, respectively. We also provide the mean values ($\blacklozenge$).
	
	The LEVSS algorithm consistently outperforms the IBOSS and OSS ones in Cases 2 and 3. Also, in Case 1, LEVSS algorithm consistently outperforms the IBOSS one, and it is slightly better than the OSS one. This happens because the structure of the selected subdata by the LEVSS and OSS algorithm are very similar, when the observations $\textbf{x}_i$'s, $i=1,2,\ldots,50$ are independent and follow a multivariate uniform distribution with all covariates independent. Moreover, MSE of the estimated slope parameters by LEVSS approach decreases as the full data size $n$ increases, even though the subdata size is fixed at $k=1000$. This indicates that LEVSS approach identifies more informative data points from the full data as the full data size increases. Also, either the covariates are unbounded or not, as $n$ increases, the MSE of the estimated slope parameters decreases fast in the LEVSS approach, as in the OSS one. Overall, LEVSS approach provide more accurate estimates for the model parameters compared with the IBOSS and OSS approaches, as one can see in Figures  \ref{fig_uni}, \ref{fig_norm} and \ref{fig_tnorm}. The MSE for intercept is immutable among the three approaches, and so the results are omitted for brevity. For the results from the IBOSS and OSS approaches, we refer the reader to Figure 1 in the supplementary material by \cite{wang2021oss}.
	
	\subsection{\label{inter} Model with interactions}
	We are interested in evaluating the approaches of IBOSS, OSS and LEVSS, considering the existence of interactions between covariates. The response data are generated, for each case of covariates in subsection \ref{section_first_order}, from the model
	$$
	\textbf{y} = \beta_0+\textbf{X}_{m}\boldsymbol{\beta}_{m}+\textbf{X}_{\gamma}\boldsymbol{\beta}_{\gamma}+\boldsymbol{\epsilon},
	$$
	where each columns of $\textbf{X}_{m}$ is a covariate, each column of $\textbf{X}_{\gamma}$ is an element-wise product of two columns in $\textbf{X}_{m}$,  $\boldsymbol{\beta}_{m}=(\beta_1,\beta_2,\ldots,\beta_p)^{\text{T}}$ is a $p$-dimensional vector of main effects and $\boldsymbol{\beta}_{\gamma}=(\beta_{p+1},\beta_{p+2},\ldots,\beta_{p(p-1)/2})^{\text{T}}$ is a $p(p-1)/2$-dimensional vector of interaction effects, and $\boldsymbol{\epsilon}$ are the error terms.
	
	We set $p=10$, $\beta_0=1$, the true value of $\boldsymbol{\beta}_{m}$ and $\boldsymbol{\beta}_{\gamma}$ being $10$ and $45$ dimensional vectors of unity, respectively, and $\sigma^2=9$. 
	
\begin{figure}[!thb]
		{\centering
			\includegraphics[width=1\textwidth]{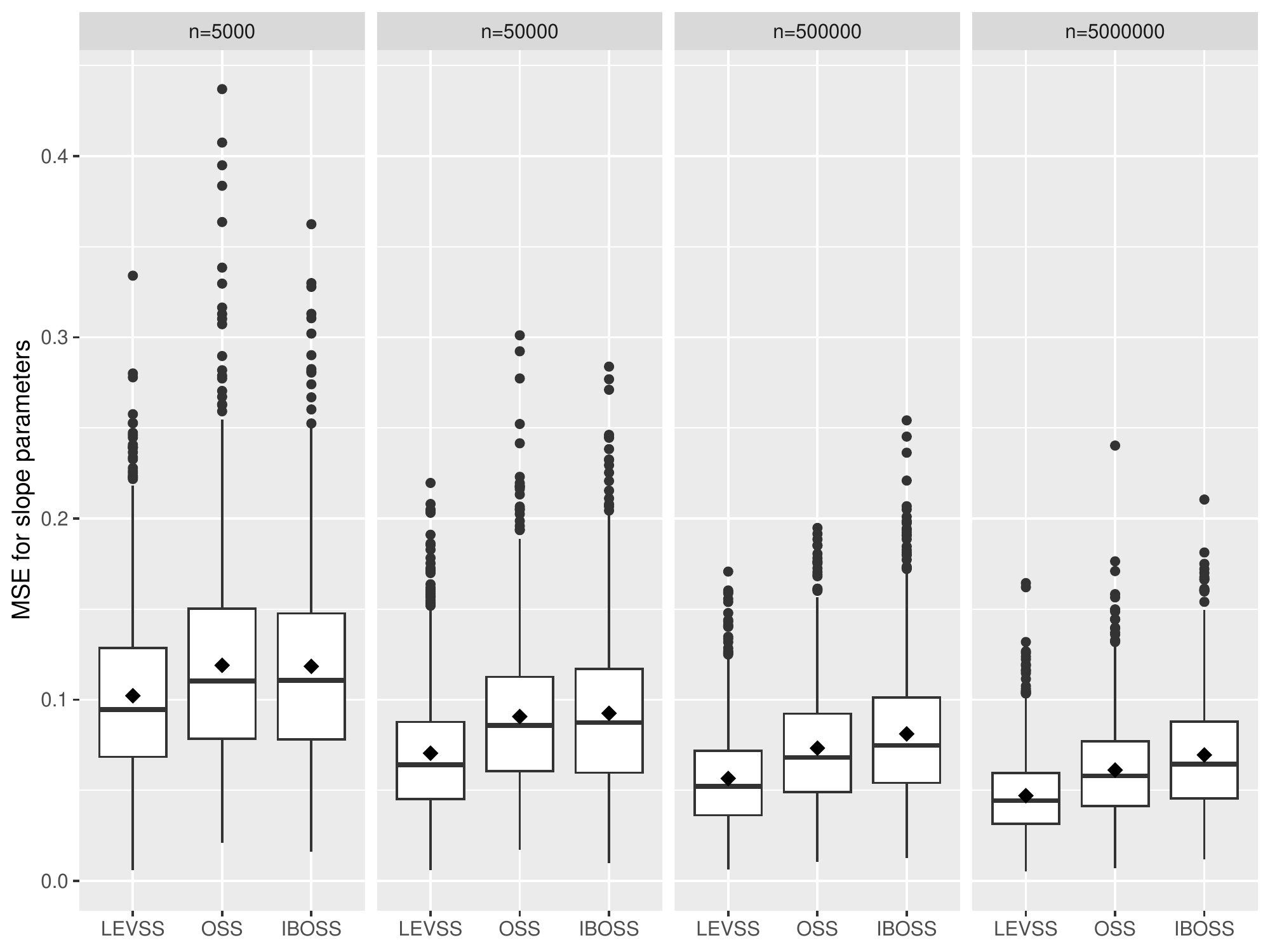}\par}
		\caption{MSE$_{\boldsymbol{\beta}_{m}}$ for the subdata selected by different approaches for the covariates of Case 3, when the full data size is $n=5\times10^3, 10^4, 10^5$ and $10^6$ and the subdata size is $k=1000$.}
		\label{fig_tnorm_me}
	\end{figure}

\begin{figure}[!thb]
		{\centering
			\includegraphics[width=1\textwidth]{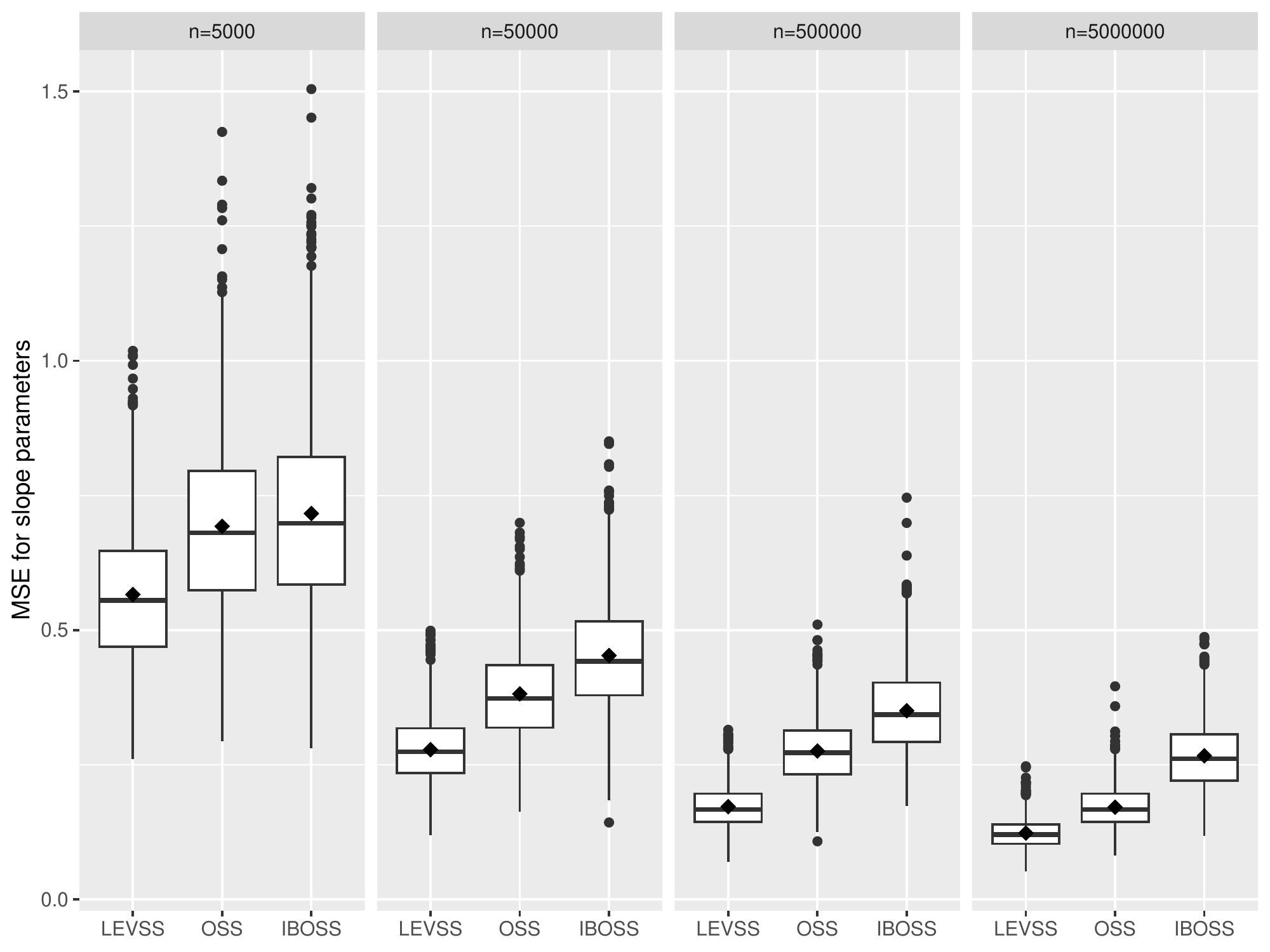}\par}
		\caption{MSE$_{\boldsymbol{\beta}_{\gamma}}$ for the subdata selected by different approaches for the covariates of Case 3, when the full data size is $n=5\times10^3, 10^4, 10^5$ and $10^6$ and the subdata size is $k=1000$.}
		\label{fig_tnorm_inter}
	\end{figure}

	The simulation is repeated $1000$ times. We calculate the MSEs of the estimated $\boldsymbol{\beta}_{m}$ and $\boldsymbol{\beta}_{\gamma}$, that is MSE$_{\boldsymbol{\beta}_{m}}$ and MSE$_{\boldsymbol{\beta}_{\gamma}}$, separately, for the subdata selected by the approaches of IBOSS, OSS and LEVSS. We investigate the cases that the full data sizes are $n=5\times10^3, 10^4, 10^5$ and $10^6$, and the subdata size is fixed at $k=1000$. Figures \ref{fig_tnorm_me} and \ref{fig_tnorm_inter} show MSE$_{\boldsymbol{\beta}_{m}}$ and MSE$_{\boldsymbol{\beta}_{\gamma}}$ for the subdata selected by different approaches for Case 3. We also provide the mean values ($\blacklozenge$). The results for Cases 1 and 2 are similar, and so they are omitted for brevity. 
	
	First, MSE$_{\boldsymbol{\beta}_{m}}$ for the subdata selected by the LEVSS approach is lower than those of the approaches of IBOSS and OSS, and so LEVSS provides more accurate estimations of main effects compared to the remaining approaches. Also, the LEVSS approach is able to identify significant interaction effects compared to the approaches of IBOSS and OSS, since MSE$_{\boldsymbol{\beta}_{m}}$ by the LEVSS approach is the smallest one. It is important to mention that the approaches of both LEVSS and OSS select data points based on covariates, while the approach of IBOSS relies on the interaction terms as well. 
	
	\subsection{\label{time} Computing time}
	We focus on the computing time of the approaches IBOSS, OSS and LEVSS for Case 1 for different full data sizes $n$, when the subdata size is equal to $k=1000$ and the number of covariates is equal to $p=50$. All computations are carried out on a PC with 3.6 GHz Intel 8-Core I7
	processor and 16GB memory. 
	
	In Table \ref{table1}, we present the mean computing times (in seconds) of the approaches IBOSS, OSS and LEVSS. 
	
	\begin{table}[!htb]
		\centering
		\setlength{\tabcolsep}{3pt}
		\begin{tabular}{ccccc}
			\toprule
			$n$ & $5\times10^3$ & $5\times10^4$ & $5\times10^5$ & $5\times10^6$ \\ \midrule
			IBOSS & 0.175 & 0.758 & 6.858 & 73.14 \\ \midrule
			OSS & 3.205 & 6.886 & 18.351 & 150.06 \\ \midrule
			LEVSS & 1.205 & 1.812 & 7.911 & 83.16  \\ \bottomrule
		\end{tabular}
		\caption{The mean execution time (in seconds) of the approaches IBOSS, OSS and LEVSS for different full data sizes $n=5\times10^3, 10^4, 10^5$ and $10^6$, when the subdata size is equal to $k=1000$ and the number of covariates is equal to $p=50$.}
		\label{table1}
	\end{table}
	
	For any full data size $n$, the algorithm of the LEVSS approach is faster than the one of the OSS approach. Also, noting that the difference in the computing time between the algorithms of the LEVSS and the IBOSS approach is very small.

	\section{Real data application}
	\label{section_real_data}
	
	In this section, we evaluate the performance of the LEVSS approach
	approach on a real data example, examining the accuracy of the ordinary least-square estimator of slope parameters in model \eqref{model1}.
	
	The dataset of the real data example consists of locations and absorbed power of wave energy converters in four real wave scenarios from the southern coast of Australia (Sydney, Adelaide, Perth and Tasmania). The full data consists of $n = 288, 000$ data
	points and contains readings of $32$ location variables and $16$ absorbed power variables, so the number of covariates in the model is $p = 48$. The response variable is the total power output
	of the farm, and in the analysis we work with its log-transformation. Further information about the dataset can be found in ``UCI Machine Learning Repository" \cite{dua2019}.
	
	We compare the performance of the algorithm of the LEVSS approach with the algorithms of the approaches of IBOSS and OSS, considering the MSE for the vector of slope parameters for each algorithm by using $100$ bootstrap samples, as in \cite{wang2019information} and \cite{wang2021oss}. Each bootstrap sample is a random sample of size $n$ from the full data using sampling with replacement. For a bootstrap sample, each algorithm is implemented to obtain the subdata and then from the selected subdata the parameters of the model are estimated. The algorithm of the LEVSS approach is implemented under different values for the threshold $T$ on the condition number, that is for $T=25$, $20$ and $15$, and when the stopping criterion on the condition number does not exist. These modifications on the algorithm of the LEVSS approach take place in the real data example, but not in the simulation experiments in Section \ref{section_simulation}, since the condition number was too small when the $k$ have already been selected, that is the space of covariates of the subdata expanded quickly to the space of covariates of the full data.
	
\begin{figure}[!thb]
		\centering
		\includegraphics[width=1\textwidth]{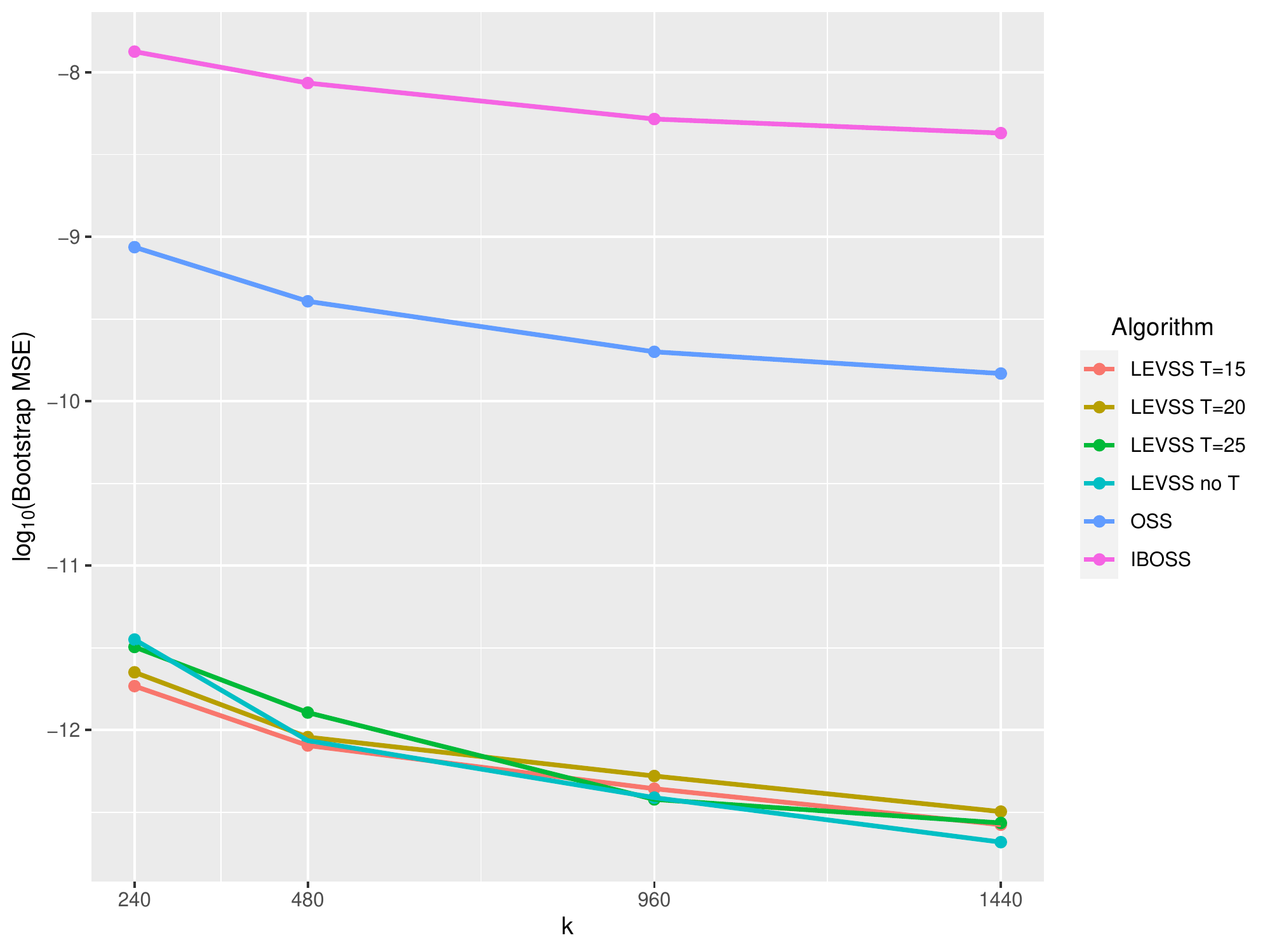}
		\caption{The bootstrap MSEs for the subdata selected by different approaches, when the subdata size is $k=5p, 10p, 20p$ and $30p$. The different values for the threshold $T$ on the condition number are equal to $25$ (LEVSS $T=25$), $20$ (LEVSS $T=20$) and $15$ (LEVSS $T=15$). Also, the algorithm of the LEVSS approach is implemented without the stopping criterion on the condition number (LEVSS no $T$).}
		\label{energy}
	\end{figure}

	Figure \ref{energy} shows the bootstrap MSEs by different approaches for $k=5p, 10p, 20p$ and $30p$.  Also, we take logarithm with base 10 of each MSE for a better presentation of the Figure \ref{energy}. Moreover, we provide the mean values ($\blacklozenge$). All modifications on the algorithm of the LEVSS approach outperform the IBOSS and OSS algorithms in minimizing the bootstrap MSEs. Also, as \cite{wang2021oss} stated, the IBOSS approach performs poorly in this real data example, because probably not all variables are important. On the other hand, one could say that the LEVSS approach approximates as close as possible the convex hull generated by the full data, and so performs very well. It is important to make a discussion on the modifications of the algorithm of the LEVSS approach. As the subdata size $k$ is getting bigger, the bootstrap MSE is the smallest one for the case that the stopping criterion on the condition number is ignored. Also, consider that the LEVSS approach acts like simple random subsampling on the full data in some sense when the $T$ is getting smaller, since then more than $k$ data points are selected. This consideration seems to make sense as the subdata size $k$ is getting larger, since for a small $k$, say $k=240$, the bootstrap MSE is the smallest for the case that the $T$ is the smallest one among others used. However, another value of $T$, which is lower that $15$, will lead to a bigger bootstrap MSE for any subdata size $k$. Moreover, we should note that the algorithm of the LEVSS approach is faster when the value of $T$ is getting larger. The fastest modification of the algorithm of the LEVSS approach is when we ignore the stopping criterion on the condition number.
	
	\section{Concluding remarks}
	\label{concluding_remarks}
	We have evaluated the algorithm of the LEVSS approach for the selection of data points in an optimal way from a big dataset, in order to be able to run regression and derive the most informative coefficients as possible. Also, the algorithm of the LEVSS approach was compared with these of the IBOSS and the OSS approaches in order to show the improvement gained.
	
	Also, the modifications of the algorithm of the LEVSS approach provide very useful information about the later. It seems that a larger value on threshold $T$, or even more the absence of the stopping criterion on the condition number, can lead to the selection of more informative data points. However, one should consider the level of multicollinearity caused when the algorithm of the LEVSS approach is applied under such considerations, since as \cite{yu2022} stated, a large value on the condition number may lead to a ill-conditioned matrix and thus cause multicollinearity.
	
	Moreover, according to the results provided in \cite{yu2023}, some model-free subsampling approaches (SPARTAN, SP) perform better than the LEVSS approach in some cases of the simulation experiments. Therefore, not only a further and a more comprehensive investigation but also the development of new methods on the accommodation of real data in the big data era is required.
	
	We need to mention that we did not optimize the \texttt{R} used in anyway, and so further time savings could be possible. 
	
	\bibliographystyle{apalike}
	\bibliography{main}
\end{document}